\documentclass{aa}

\usepackage{times}
\usepackage{graphics}

\begin{document}
   \thesaurus{11(05.01.1;08.02.3)}
   \title{Indications on the binary nature of individual stars
          derived from a comparison of their HIPPARCOS proper motions
          with ground-based data}
   \subtitle{I. Basic principles}

   \author{R. Wielen
   \and C. Dettbarn
   \and H. Jahrei{\ss}
   \and H. Lenhardt
   \and H. Schwan}

   \offprints{R. Wielen (wielen@ari.uni-heidelberg.de)}

   \institute{Astronomisches Rechen-Institut, Moenchhofstrasse 12-14,
   D-69120 Heidelberg, Germany}

   \date{Received 2 February 1999 / Accepted 26 March 1999}

\authorrunning{R. Wielen et al.}
\titlerunning
{Indications for delta-mu binaries and single-star candidates.
I. Basic principles}

\maketitle

\begin{abstract}

We present a method which provides some information on the possible binary
nature of an apparently single star. The method compares the instantaneously
measured HIPPARCOS proper motion with the long-term averaged, ground-based
proper motion or with the proper motion derived from old ground-based positions
and the HIPPARCOS position. Good sources for such ground-based data are
the FK5 and the GC.

If the proper-motion difference $\Delta\mu$
is statistically significant with respect to
its measuring error, the object is very probably a double star. We call then
the object a 'delta-mu binary'. If the proper-motion
difference is insignificant and if no other information on a binary nature of
the object is available, we call such a star a 'single-star candidate'.

We propose a quantitative test for the significance of the observed
proper-motion difference. The sensitivity of our method is high: For nearby
stars at a distance of 10 pc, the measuring accuracy
of the proper-motion difference, expressed as a velocity,
is of the order of 50 m/s (basic FK5 stars) or 80 m/s (GC stars). At 100 pc,
the mean error of the two-dimensional difference is still 0.5 km/s or 0.8 km/s.

For the FK5 stars, we provide indications on the probable period of the
$\Delta\mu$ binaries.
If we adopt an orbital period and a mass-luminosity relation, we can
use the observed velocity difference to estimate the separation and the
magnitude difference between the two components of the binary.

The present paper concentrates mainly on the basic principles of the method,
but it provides also a few examples of delta-mu binaries and of single-star
candidates for illustration: $\gamma$ UMa, $\varepsilon$ Eri, $\iota$ Vir, 47
UMa, $\delta$ Pav.


\end{abstract}

\section{Introduction}

For many purposes it is important to know whether an object is a single star or
a binary. There are many conventional methods to detect the binary nature of an
object: direct imaging methods (from visual inspection to speckle
observations), photometric methods (eclipsing binaries), and detection of
orbital motions (spectroscopic and astrometric binaries). The high measuring
accuracy of the ESA Astrometry Satellite HIPPARCOS
has added some new methods for
detecting binaries, which lead to the component (C) solutions,
acceleration (G) solutions, variability-induced movers (V solutions), and
stochastic (X) solutions, in addition to the classical orbital (O) solutions
(ESA, 1997).

We discuss here an additional method to detect the binary nature of an
apparently single star: The new method is based on the comparison of the
quasi-instantaneously measured HIPPARCOS proper motion with a time-averaged,
long-term proper motion derived either from ground-based observations alone or
from a combination of old ground-based positions with the HIPPARCOS position.
The basic idea is illustrated in Fig. 1. For a truly single star, the proper
motion measured within a short interval of time should agree, within the
measuring accuracy, with the proper motion derived from a very long interval.
This is in general not true for a binary: Due to the wavy orbital motion of the
photo-center of an unresolved astrometric binary, the instantaneously measured
proper motion of the object can differ significantly from a long-term proper
motion (ideally the motion of its center-of-mass). We have called the
difference between the two proper motions the `cosmic error' of the
instantaneous proper motion (Wielen 1995a, b, 1997, Wielen et al. 1997).

If we detect for an object an individual cosmic error which is significantly
larger than the measuring error, then this object is very probably a binary. We
call these objects `delta-mu binaries' ($\Delta\mu$ binaries) because of the
proper-motion difference $\Delta\mu$
which has led to the detection of their double-star
nature. In the opposite case, i.e. if the individual cosmic error is well
within the expectation provided by the measuring errors, then the object is
either actually a single star or the orbital motion of the photo-center was too
small to be detected with the given measuring accuracy. We call such an object
a `single-star candidate', if there is nothing known to us that actually
indicates a binary nature of this object (either from ground-based observations
or from HIPPARCOS data). There remains, of course, the third possibility: The
cosmic error is neither large enough for qualifying the star as a $\Delta\mu$
binary nor small enough for a single-star candidate.

\begin{figure}[t]
\resizebox{\hsize}{!}{\includegraphics{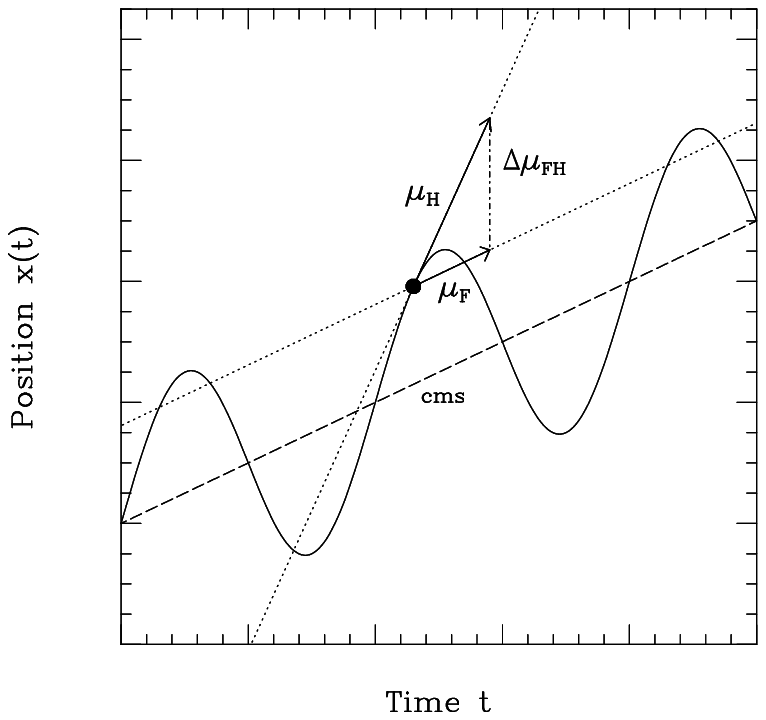}}
\caption[]
{The wavy motion of an astrometric binary leads to an observable difference
$\Delta\mu_{FH}$ between the instantaneously measured HIPPARCOS proper motion
$\mu_H$ and the mean proper motion $\mu_F$ of the star. We assume in this
example that the orbital period of the binary is of medium length (e.g. 30
years), so that the proper motion $\mu_F$, obtained from ground-based data
(e.g. from the FK5),
is essentially equal to the proper motion of the center-of-mass
(cms) of the binary.}
\end{figure}

\section{Proper-motion differences}

Our method is based on the comparison of an `instantaneously' measured proper
motion with at least one long-term proper motion. Short-term proper motions,
$\mu_H$, derived from observations over a period of about 3 years around 1991,
are provided by HIPPARCOS (ESA 1997).

There are various possibilities for getting long-term proper motions. Let us
start with the FK5 (Fricke et al. 1988, 1991). This catalogue of fundamental
stars provides rather accurate long-term proper motions, $\mu_F$, derived from
ground-based observations which cover periods of up to more than 200 years. The
FK5 gives also mean stellar positions, $x_F (T_F)$. We use the notation $x$ for
either the right ascension $\alpha$ or the declination $\delta$ of the star.
The individual `central' epoch  $T_F$
is chosen such that $x_F$ and $\mu_F$ are not
correlated. From the HIPPARCOS position, $x_H (T_H)$, with $T_H \sim 1991.25$,
and the FK5 position, we can derive a second long-term proper motion $\mu_0$:
\begin{equation}
\mu_0 = \frac{x_H (T_H) - x_F (T_F)}{T_H - T_F} \,\, .            
\end{equation}
The epoch difference $T_H - T_F$ is typically of the order of 40 years. Since
$x_F (T_F)$ and $\mu_F$ are uncorrelated, this is also true for $\mu_0$ and
$\mu_F$. The same is true for the pair $\mu_0$ and $\mu_H$, if we use the
individual central epochs $T_H$ for each star and for the two directions
$\alpha$ and $\delta$.

The values $\mu_F$ and $x_F$ should not be taken directly from the FK5, since
the catalogue values are affected by systematic errors. Instead we have first
to reduce the positions and proper motions given in the FK5 to the HIPPARCOS
system (see e.g. Wielen 1997, Wielen et al. 1998, 1999,
based on the method of
Bien et al. (1978)). In the case of the FK5
we can now form the differences for the three pairs
of the three proper motions, separately for
each coordinate component $\alpha\,\cos\,\delta = \alpha_{\ast}$ and $\delta$.
For $\alpha_{\ast}$, we get
\begin{eqnarray}
\Delta\mu_{FH, \alpha\ast} & = & \mu_{F, \alpha\ast} - \mu_{H, \alpha\ast} \,\,
,\\                                                                    
\Delta\mu_{0H, \alpha\ast} & = & \mu_{0, \alpha\ast} - \mu_{H, \alpha\ast} \,\,
,\\                                                                    
\Delta\mu_{0F, \alpha\ast} & = & \mu_{0, \alpha\ast} - \mu_{F, \alpha\ast} \,\,
,                                                                      
\end{eqnarray}
and three similar equations for $\delta$. The differences in the two directions
$\alpha$ and $\delta$
can be added together to a `total' difference:
\begin{equation}
\Delta\mu_{FH, tot}
= \Big ((\Delta\mu_{FH, \alpha\ast} )^2
+ (\Delta\mu_{FH, \delta})^2 \Big )^{1/2} \,\, ,                  
\end{equation}
and similar equations for the pairs $0H$ and $0F$. Because the central epochs
$T$ are usually different for $\alpha$ and $\delta$, the difference
$\Delta\mu_{tot}$ is not strictly a proper-motion difference for a certain
epoch. This complicates slightly the physical interpretation of
$\Delta\mu_{tot}$. But any choice of a common epoch for $\alpha$ and $\delta$
would introduce correlations which would disturb our statistical considerations
in Sect. 3.

Already for the FK5, the proper motion $\mu_0$ is usually more accurately
measured than $\mu_F$. This is due to the rather good accuracy of old
ground-based observations coupled with their high epoch difference with respect
to HIPPARCOS. An important compilation catalogue of such older observations is
the GC (Boss et al. 1937). The GC contains many more stars (33\,342) than
the FK5 (4\,652), albeit with lower accuracy. Unfortunately the proper
motions, $\mu_{GC}$, given in the GC have such large measuring errors that they
can not be used for our purpose in most cases. However, the proper motion
$\mu_{0(GC)}$, based on the GC position $x_{GC} (T_{GC})$ at $T_{GC} \sim
1900$,
\begin{equation}
\mu_{0(GC)} = \frac{x_H (T_H) - x_{GC} (T_{GC})}{T_H - T_{GC}} \,\, ,    
\end{equation}
is usually accurate enough for being used in our method. Since we have now only
two proper motions, $\mu_{0(GC)}$ and $\mu_H$, for a comparison, our formerly
three  pairs of proper-motion differences are reduced to one difference in the
case of the GC:
\begin{eqnarray}
\Delta\mu_{0(GC) H, \alpha\ast} & = & \mu_{0(GC), \alpha\ast} - \mu_{H,
\alpha\ast} \,\, ,\\                                                   
\Delta\mu_{0(GC) H, \delta} & = & \mu_{0(GC), \delta} - \mu_{H, \delta } \,\, ,
\end{eqnarray}                                                         
and
\begin{equation}
\Delta\mu_{0(GC) H, tot}  =  \Big ((\Delta\mu_{0(GC) H, \alpha\ast})^2
                          + (\Delta\mu_{0(GC) H,
                                        \delta})^2 \Big )^{1/2}  .  
\end{equation}
Most of the FK5 stars are also contained in the GC. For these common stars, a
comparison of $\Delta\mu_{0H}$ and $\Delta\mu_{0(GC) H}$ provides a partial
consistency check. One should remember, however, that the FK5 and the GC often
use the same old observations and differ only in the detailed treatment of
these common data.

Up to now, we have implicitely assumed that the coordinates $\alpha_{\ast}$ and
$\delta$ of a star change linearly in time. In real applications, the effects
of sphericity and the foreshortening effect have to be taken into account. This
must be done, however, already for the determination of the systematic
differences between the FK5 (or the GC) and HIPPARCOS. If we use later always
the HIPPARCOS results as reference values, i.e. $x_F - x_H$ instead of $x_F$
and $\mu_F - \mu_H$ instead of $\mu_F$, then the non-linear effects are already
accounted for to a high degree of approximation.

\section{Statistical significance}

After having derived in Sect. 2 the proper-motion difference for a given
star, we have now to decide whether or not this difference is statistically
significant with respect to its measuring error.

The measuring error $\varepsilon_{\Delta\mu, FH, \alpha\ast}$ of
$\Delta\mu_{FH, \alpha\ast}$ is given by
\begin{equation}
\varepsilon^2_{\Delta\mu, FH, \alpha\ast} =
\varepsilon^2_{\mu, F, \alpha\ast} + \varepsilon^2_{\mu, H, \alpha\ast} \,\, .
\end{equation}                                                       
$\varepsilon_{\mu, H, \alpha\ast}$ is the measuring error of $\mu_{H,
\alpha\ast}$, given in the HIPPARCOS Catalogue. The measuring error
$\varepsilon_{\mu, F, \alpha\ast}$ of $\mu_{F, \alpha\ast}$ consists of two
parts:
\begin{equation}
\varepsilon^2_{\mu, F, \alpha\ast} = \varepsilon^2_{\mu, F, \alpha\ast, ind} +
\varepsilon^2_{\mu, F, \alpha\ast, sys} \,\, .                        
\end{equation}
The first part is the random (`individual') mean error of $\mu_{F,
\alpha\ast}$, provided by the FK5. The second part describes the uncertainty of
the reduction of the FK5 system of proper motions to the HIPPARCOS system for
the star under consideration. We emphasize that this is not the systematic
difference of the FK5 itself, but only the mean error of the determination of
this systematic difference.

The mean error $\varepsilon_{\mu, 0, \alpha\ast}$ of $\mu_{0, \alpha\ast}$ is
obtained from
\begin{equation}
\varepsilon^2_{\mu, 0, \alpha\ast} = \frac{
  \varepsilon^2_{x, H, \alpha\ast}
+ \varepsilon^2_{x, F, \alpha\ast, ind}
+ \varepsilon^2_{x, F, \alpha\ast, sys}
}{(T_{H, \alpha\ast} - T_{F, \alpha\ast})^2} \,\,  .         
\end{equation}
$\varepsilon_{x, H, \alpha\ast}$ is the measuring error of the HIPPARCOS
position $x_H (T_H)$ in $\alpha_\ast$, while
$\varepsilon_{x, F, \alpha\ast, ind}$
is the random error of $x_F (T_F)$,
and $\varepsilon_{x, F, \alpha\ast, sys}$ is the
uncertainty in the reduction of the FK5 system of positions to the HIPPARCOS
system for this star. The measuring errors of the proper-motion differences for
the pairs $0H$ and $0F$ are then given by
\begin{eqnarray}
\varepsilon^2_{\Delta\mu, 0H, \alpha\ast} & = & \varepsilon^2_{\mu, 0,
\alpha\ast}  +  \varepsilon^2_{\mu, H, \alpha\ast} \,\, ,\\          
\varepsilon^2_{\Delta\mu, 0F, \alpha\ast} & = & \varepsilon^2_{\mu, 0,
\alpha\ast}  +  \varepsilon^2_{\mu, F, \alpha\ast} \,\, .            
\end{eqnarray}
The corresponding equations for the coordinate $\delta$ have the same
form: $\alpha_\ast$ is just replaced by $\delta$. Equations (10)-(14) are
applicable when using the FK5. Equations (12) and (13) can be easily adapted to
the case of the GC by replacing $F$ by $GC$ and $0$ by $0(GC)$.

We have also to take into account that the components $\mu_{H, \alpha\ast}$ and
$\mu_{H, \delta}$ of the HIPPARCOS proper motions are correlated. The
corresponding correlation coefficient $\rho_{\mu\alpha\ast, \mu\delta}$ is
given in the HIPPARCOS Catalogue. The components of the ground-based proper
motions, $\mu_F$ or $\mu_{GC}$, are not correlated with $\mu_H$. All the
correlations of the components of $\mu_0$ with other quantities are neglected
here, since they are
usually very small, because the measuring error of the ground-based position
$x_F (T_F)$ is always much larger than that of the HIPPARCOS position $x_H
(T_H)$. For the same reason, we neglect the cross-correlations between $x_{H,
\alpha\ast}$ and $\mu_{H, \delta}$, and between $x_{H, \delta}$ and $\mu_{H,
\alpha\ast}$. From the correlation coefficient $\rho_{\mu\alpha\ast,
\mu\delta}$, we can derive the covariance $\gamma$ which is the same for
$\mu_H$, $\Delta\mu_{0H}$, and $\Delta\mu_{FH}$:
\begin{equation}
\gamma = \rho_{\mu\alpha\ast, \mu\delta} \,\, \varepsilon_{\mu, H,
\alpha\ast} \,\, \varepsilon_{\mu, H, \delta} \,\, .                  
\end{equation}

\begin{figure}[t]
\resizebox{\hsize}{!}{\includegraphics{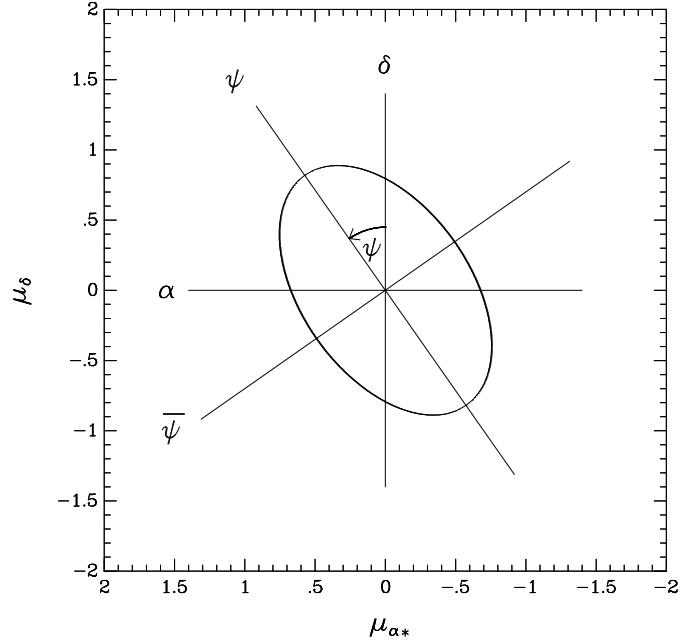}}
\caption[]
{The error ellipsoid of the measuring errors of $\Delta\mu$ is tilted with
respect to the equatorial system ($\delta,\alpha$) by an angle $\psi$.
The major
axis of the error ellipsoid points in the direction $\psi$, the minor axis
towards $\overline \psi$.}
\end{figure}

We shall now discuss the statistical significance of the proper motion
difference $\Delta\mu_{0H}$. The result for this pair of proper motions is,
however, directly adaptable for the two other differences, $\Delta\mu_{FH}$ and
$\Delta\mu_{0F}$.

We assume that the measuring errors in $\Delta\mu_{0H, \alpha\ast}$ and
$\Delta\mu_{0H, \delta}$ follow Gaussian distributions with mean zero and
dispersions $\varepsilon_{\Delta\mu, 0H, \alpha\ast}$ and
$\varepsilon_{\Delta\mu, 0H, \delta}$. However, if $\gamma \neq 0$, the
directions of $\alpha$ and $\delta$ are generally not the principal axes
of the error ellipsoid. Instead, these principle axes are rotated with
respect to the equatorial system by an angle $\psi$ (see Fig. 2).
The angle $\psi$ is derived from
\begin{eqnarray}
sin \, 2 \psi & = & 2 \gamma\,/\,k \, \, ,\\                      
cos \, 2 \psi & = & - \left(\varepsilon^2_{\Delta\mu, 0H, \alpha\ast} -
                      \varepsilon^2_{\Delta\mu, 0H, \delta}\right)\,/\,k \,\, ,
\end{eqnarray}                                                         
with the auxiliary quantity
\begin{equation}
k = + \left(\left(\varepsilon^2_{\Delta\mu, 0H, \alpha\ast} -
\varepsilon^2_{\Delta\mu,
0H, \delta}\right)^2 + 4 \, \gamma^2\right)^{1/2} \,\, .               
\end{equation}
The angle $\psi$ is counted from North towards the East, like a position angle.
The dispersions along the principle axes of the error ellipsoid are then
given by
\begin{eqnarray}
\varepsilon^2_{\Delta\mu, 0H, \psi} & = & \frac{1}{2} \,
\left(\varepsilon^2_{\Delta\mu, 0H, \alpha\ast} + \varepsilon^2_{\Delta\mu, 0H,
\delta} + k\right) \,\, ,\\                                            
\varepsilon^2_{\Delta\mu, 0H, {\overline \psi}} & = & \frac{1}{2} \,
\left(\varepsilon^2_{\Delta\mu, 0H, \alpha\ast} + \varepsilon^2_{\Delta\mu, 0H,
\delta} - k\right) \,\, .                                              
\end{eqnarray}
The components of the observed proper-motion difference $\Delta\mu_{0H}$ in the
system of the principle axes of the error ellipsoid
(directions $\psi$ and $\overline \psi$)
are
\begin{eqnarray}
\Delta\mu_{0H, \psi} & = & + \Delta\mu_{0H, \alpha\ast} \,
\sin\,\psi + \Delta\mu_{0H, \delta} \, \cos\,\psi \,\, ,\\   
\Delta\mu_{0H, {\overline \psi}} & = & + \Delta\mu_{0H, \alpha\ast} \,
\cos\,\psi - \Delta\mu_{0H, \delta} \,\sin\,\psi \,\, .     
\end{eqnarray}

\begin{figure}[t]
\resizebox{\hsize}{!}{\includegraphics{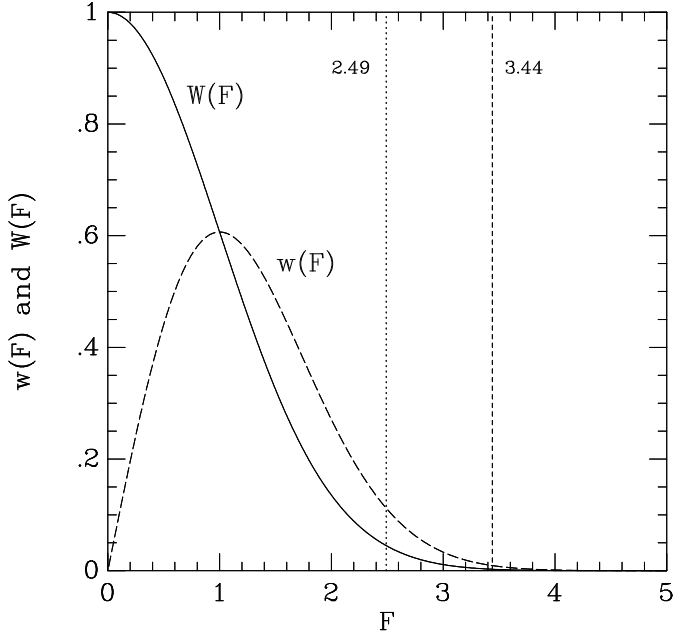}}
\caption[]
{The function $W(F)$, given by Eq. (24),
describes the probability to find by
chance an observed value of the test parameter larger than
$F$. The differential
probability $w(F)$ is given by Eq. (25).
The two adopted critical values,
$F > 3.44$ for $\Delta\mu$ binaries and
$F < 2.49$ for single-star candidates, are
indicated.}
\end{figure}

Instead of discussing now the statistical significance of $\Delta\mu_{0H,
\psi}$ and $\Delta\mu_{0H, {\overline \psi}}$ separately as two linear
problems, it is more suitable to discuss the significance of the vector
$\Delta\mu_{0H}$ as a two-dimensional problem. For this purpose, we define the
test parameter $F_{0H}$ by
\begin{equation}
F^2_{0H} = \bigg( \frac{\Delta\mu_{0H, \psi}}{\varepsilon_{\Delta\mu, 0H,
\psi}} \bigg)^2 +
\bigg( \frac{\Delta\mu_{0H, {\overline \psi}}}{\varepsilon_{\Delta\mu, 0H,
{\overline \psi}}} \bigg)^2 \,\, .
\end{equation}
In the 'isotropic' case of the measuring errors (i.e. for
$\varepsilon_{\Delta\mu,0H,\alpha\ast}$ =
$\varepsilon_{\Delta\mu,0H,\delta}$ = $\varepsilon_{\Delta\mu,0H,1\rm D}$ and
$\rho_{\mu\alpha\ast,\mu\delta}$ = 0),
$F$ would simply be the ratio between the
proper-motion difference $\Delta\mu_{0H,tot}$ and its measuring error
$\varepsilon_{\Delta\mu,0H,1\rm D}$.

If the star is not a binary, then the uncorrelated
variables $\Delta\mu_{0H, \psi}$ and
$\Delta\mu_{0H, \overline \psi}$
are expected to follow normal distributions with mean
zero and dispersions according to Eqs. (19) and (20). In this case, the
probability $W (F)$ to find by chance a value of $F_{0H}$ which is equal to or
larger than the observed value (given by Eq. (23)), is
\begin{equation}
W (F) = e^{-F^2_{0H}/2} \,\, .                                         
\end{equation}
The differential probability $w (F)\,dF$ to find $F_{0H}$ between $F$ and $F +
dF$ is given by
\begin{equation}
w (F) = - \frac{dW (F)}{dF} = F_{0H} \,\, e^{-F^2_{0H}/2} \,\, . 
\end{equation}
The function $w (F)$ is plotted in Fig. 3. For small $F,\,w (F)$ increases
linearly with $F$. The function $w (F)$ reaches a maximum at $F = 1$, and
declines rapidly for larger values of $F$.

We conclude from the run of the function $W(F)$
that a high observed value of $F_{0H}$ is a strong
indication for the binary nature of the object under consideration. Which
minimal value $F_{lim, b}$ for $F_{0H}$ should be used for our $\Delta\mu$
binaries\,? We propose to call those objects $\Delta\mu$
{\it {\underline b}inaries}
for which
\begin{equation}
F_{0H} > F_{lim, b} = 3.44                                              
\end{equation}
holds. This gives the same level of significance,
\begin{equation}
W(3.44) = 0.0027 \,\, ,                                               
\end{equation}
as the often used, two-sided 3$\sigma$ criterion for a one-dimensional normal
distribution. It means that among 10\,000 truly single stars, only 27 of them
would be wrongly classified by us as $\Delta\mu$ binaries. Our proposed value
of 3.44 has also been adopted for the $G$ solutions in the HIPPARCOS Catalogue.

While a large observed value of $F_{0H}$ hints strongly to a binary nature of
the object, a small value of $F_{0H}$ makes it rather probable that the star is
either single or that its orbital motion is below the level set by the
measuring accuracy of $\Delta\mu_{0H}$. We propose to call objects with
\begin{equation}
F_{0H} < F_{lim, s} = 2.49                                             
\end{equation}
{\it {\underline s}ingle-star candidates}.
This limit corresponds to a $2 \sigma$ criterion,
\begin{equation}
W(2.49) = 0.0456  \,\, .                                              
\end{equation}
From 20 truly single stars, 19 of them would be correctly classified as
single-star candidates. One of them
would be wrongly dismissed. In general, it is
impossible to say how many actual binaries are wrongly classified as
single-star candidates. This depends on the measuring accuracy of the
proper-motion difference and on the distribution function of the orbital
velocities of the binaries. We should therefore strongly emphasize the word
`candidate' in our term `single-star candidate'.

In the case of the GC, we have only one meaningful test parameter, namely
$F_{0 (GC) H}$ . For the FK5, however, three test parameters are available:
$F_{FH}, \, F_{0H}$, and $F_{0F}$ . They are, of course, not independent, since
the proper-motion differences are related by $\Delta\mu_{FH} = \Delta\mu_{0H}
- \Delta\mu_{0F}$ . While there are various possibilities to combine the three
proper-motion differences into a single test parameter, we propose to call an
object a `$\Delta\mu$ binary'
if {\em at least one} $F$ value, either
$F_{FH}$ or $F_{0H}$ or $F_{0F}$, is larger than $F_{lim, b}$ . For a
`single-star candidate', {\em all} the $F$ values should be smaller than
$F_{lim, s}$ . The reasons for this choice will become clearer by the
discussion in the next section.

\section{Which types of binaries can be detected\,?}

The types of binaries which can be detected depend, of course, on the nature
and quality of the available astrometric data. We concentrate here on the cases
FK5\,+\,HIPPARCOS and GC\,+\,HIPPARCOS.

Since HIPPARCOS proper motions and positions are already averages over about 3
years of observations, short-period binaries with orbital periods $P$ below 3
years should not be detectable by our method.

For binaries with medium periods of a few decades, say $P \sim 30$ years, the
HIPPARCOS proper motion $\mu_H$ is essentially an instantaneous value, while
the other proper motions $(\mu_F, \mu_0, \mu_{0 (GC)})$ can be considered as
mean proper motions, close to the center-of-mass motion of the binary. Such
double stars can be detected by their large values of $\Delta\mu_{FH}$ and
$\Delta\mu_{0H}$ which fully contain the orbital motion of the binary. The
difference $\Delta\mu_{0F}$ between the two {\it mean}
proper motions $\mu_0$ and
$\mu_F$ should be small for such medium periods.

Double stars with long periods, say with $P \sim 1000$ years, can also be
detected by our method, if the measuring accuracy is high enough with respect
to the orbital motion of the photo-center. This is sometimes the case for
nearby objects. For long-period binaries, all the three proper motions are
essentially `instantaneously' measured values: $\mu_H$ at $T_H, \, \mu_F$ at
about $T_F$, and $\mu_0$ at about $T_0 = (T_H + T_F)/2$. The exact epochs of
$\mu_F$ and $\mu_0$ are uncertain because the values $\mu_F$ and $x_F (T_F)$
are derived from many ground-based catalogues which are spread over a long
interval of time, e.g. more than 200 years in the case of the basic FK5. On the
other hand, in these determinations the more recent catalogues have entered
with much higher weights than the old catalogues.

For the long-period binaries,
the photo-center moves approximately on a curve of second order in time. Using
$T_H$ as the reference epoch in the Taylor series, we may write:
\begin{equation}
x (T) = x_H(T_H) + \mu_H (T_H) \, (T - T_H) + \frac{1}{2} \, g \, (T -
T_H)^2  , \,                                                             
\end{equation}
and
\begin{equation}
\mu (T) = \mu_H (T_H) + g \, (T - T_H) \,\, ,                            
\end{equation}
where $g$ is the (constant) acceleration of the photo-center
in this coordinate.
We obtain then
\begin{equation}
\Delta\mu_{FH} = \mu_F (T_F) - \mu_H (T_H) = - g \, (T_H - T_F) \,\, .   
\end{equation}
Since
\begin{eqnarray}
\mu_0 (T_0) & = & \frac{x_H(T_H) - x_F(T_F)}{T_H - T_F}\nonumber\\
            & = & \mu_H (T_H) - \frac{1}{2} \, g \, (T_H - T_F) \,\, , 
\end{eqnarray}
we find
\begin{equation}
\Delta\mu_{0H} = \mu_0 (T_0) - \mu_H (T_H)
= - \frac{1}{2} \, g \, (T_H - T_F) \,\, ,                             
\end{equation}
and
\begin{equation}
\Delta\mu_{0F} = \mu_0 (T_0) - \mu_F (T_F)
= + \frac{1}{2} \, g \, (T_H - T_F) \,\, .                             
\end{equation}
Hence
\begin{equation}
\Delta\mu_{FH} = 2 \, \Delta\mu_{0H} = - 2 \, \Delta\mu_{0F} \,\, .      
\end{equation}
This means that $\Delta\mu_{FH}$ is most sensitive to the binary nature of the
object for long-period double stars.

Can we decide on the basis of our observed proper-motion differences whether
the binary has a medium period or a long one\,? For the  GC  this is
impossible, since we have only one quantity, $\Delta\mu_{0 (GC) H}$, available.
For the FK5, the situation is better. A long-period binary should fulfill
approximately the relation between the three proper-motion differences
according to Eq. (36), both in $\alpha_\ast$ and $\delta$. For a binary of
medium period and with a small cosmic error $c_x$ in the HIPPARCOS position
$x_H (T_H)$, we expect a small value of $\Delta\mu_{0F}$, so that
$\Delta\mu_{0H} \sim \Delta\mu_{FH}$. For some binaries of medium periods, the
cosmic error $c_x$ may not be negligible with respect to the measuring error
$\varepsilon_{x, F}$ of $x_F (T_F)$. Assuming that $\mu_F$ and $x_F (T_F)$ are
long-term, mean quantities, we can derive the cosmic error $c_x$ by
\begin{equation}
c_x = \Big ( (\Delta\mu_{0F})^2 - \varepsilon_{\Delta\mu,0F}^2 \Big )^{1/2}
(T_H - T_F) \,\, .                                                     
\end{equation}

In some cases, our method seems to detect also short-period binaries,
with $P \sim$ 1\,-\,3 years.
Due to the finite number and the sometimes uneven
distribution in time of the HIPPARCOS observations, the HIPPARCOS proper motion
$\mu_H$ of such short-period binaries can  deviate from the mean proper motion
(e.g. characterized by $\mu_F$) by a significant amount, even if derived from
observations spread over 3 years.

\section{The sensitivity of the method}

The sensitivity of our method is primarily determined by the astrometric
accuracy, measured in milliarcsec (mas)/year. For astrophysical considerations
it is more appropriate to translate the proper-motion difference $\Delta\mu$
into a velocity difference $\Delta\upsilon$, measured in km/s:
\begin{equation}
\Delta\upsilon \, [{\rm km/s}] = 4.74 \, \Delta\mu \, [{\rm mas/year}]\,/\,p \,
[{\rm mas}] \,\,,
\end{equation}                                                         
where $p$ is the parallax of the star. If the HIPPARCOS parallax is
statistically significant (e.g. $p > 3 \, \varepsilon_p$), we use this value.
For the remaining distant stars,
photometric or spectroscopic distances should be
preferred.

In the following general discussion, we neglect for simplicity the anisotropy
of the measuring accuracy: We replace the measuring errors of the proper
motions in $\alpha_\ast$ and $\delta$ by a common value $\varepsilon_{\mu,
1D}$:
\begin{equation}
\varepsilon^2_{\mu, 1{\rm D}} = \frac{1}{2} \, \left(\varepsilon^2_{\mu,
\alpha\ast}
+ \varepsilon^2_{\mu, \delta}\right) \,\, .                           
\end{equation}
The index 1D indicates that this rms value is the mean measuring error in
{\em one} {\em d}irection. Any correlation coefficients are also neglected in
this section.

\begin{table}[th]
\caption{
Error budget of the proper-motion differences
of 847 stars from the basic FK5
}
\begin{center}
\begin{tabular}{lccc}
\hline\\[-2.0ex]
& & \multicolumn{2}{c}{[km/s]}\\[-2ex]
Quantity & [mas/year]\\[-1ex]
& & at $r$ = 10 pc & at $r$ = 100 pc\\[0.7ex]\hline\\[-2.2ex]
$\varepsilon_{\mu, H, 1{\rm D}}$ & 0.67 & 0.032 & 0.32\\
$\varepsilon_{\mu, F, 1{\rm D}}$ & 0.84 & 0.040 & 0.40\\
$\varepsilon_{\mu, 0, 1{\rm D}}$ & 0.60 & 0.028 & 0.28\\[2ex]
$\varepsilon_{\Delta\mu, FH, 1{\rm D}}$ & 1.07 & 0.051 & 0.51\\
$\varepsilon_{\Delta\mu, 0F, 1{\rm D}}$ & 1.03 & 0.049 & 0.49\\
$\varepsilon_{\Delta\mu, 0H, 1{\rm D}}$ & 0.90 & 0.043 & 0.43\\[2ex]
3.44 $\varepsilon_{\Delta\mu, FH, 1{\rm D}}$ & 3.68 & 0.174 & 1.74\\
3.44 $\varepsilon_{\Delta\mu, 0F, 1{\rm D}}$ & 3.54 & 0.168 & 1.68\\
3.44 $\varepsilon_{\Delta\mu, 0H, 1{\rm D}}$ & 3.10 & 0.147 &
1.47\\[0.5ex]\hline
\end{tabular}
\end{center}
\end{table}

\begin{table}[th]
\caption{
Error budget of the proper-motion differences
of 11\,773 stars from the GC
}
\begin{center}
\begin{tabular}{lccc}
\hline\\[-2.0ex]
& & \multicolumn{2}{c}{[km/s]}\\[-2ex]
Quantity & [mas/year]\\[-1ex]
& & at $r$ = 10 pc & at $r$ = 100 pc\\[0.7ex]\hline\\[-2.2ex]
$\varepsilon_{\mu, H, 1{\rm D}}$ & 0.75 & 0.036 & 0.36\\
$\varepsilon_{\mu, 0\,(GC), 1{\rm D}}$ & 1.43 & 0.068 & 0.68\\[2ex]
$\varepsilon_{\Delta\mu, 0\,(GC)\,H, 1{\rm D}}$ & 1.62 & 0.077 & 0.77\\[2ex]
3.44 $\varepsilon_{\Delta\mu, 0\,(GC)\,H, 1{\rm D}}$ & 5.57 & 0.264 & 2.64\\
[0.5ex]\hline
\end{tabular}
\end{center}
\end{table}

In Tables 1 and 2 we present the error budget
for 847 basic FK5 stars and for
11\,773 GC stars. The mean errors are rms averages over the individual mean
errors
of these stars. The error budget of the stars in the FK5 extension lies between
that of the basic FK5 and the GC. The mean errors of $\mu_F$ and $\mu_0$
contain the uncertainty in the transformation of the FK5 or GC system to the
HIPPARCOS system.

We have selected those stars for which the HIPPARCOS Catalogue gives (linear)
standard solutions. In other words, these stars are not contained
in the 'Double and Multiple Star Annex (DMSA)' of the HIPPARCOS Catalogue,
i.e. they have no C, G, O, V, or X
solutions.
In addition we have excluded stars which are known to be binaries from
ground-based measurements.
For the GC, we have furthermore excluded stars with large measuring errors in
$\Delta\mu_{0(GC)H}$.

The best measured objects are the basic FK5 stars (Table 1). The sensitivity
of
our method, described by $\varepsilon_{\Delta\mu, 0H, 1{\rm D}}$, is about 0.90
mas/year. This corresponds to a mean measuring error for the velocity
difference of 0.043 km/s for a nearby object at a distance $r = 10$ pc from
the Sun. The accuracy of 43 m/s is much better than
the accuracy of conventional measurements of radial velocities and comes close
to the accuracy of the best modern radial velocities. With respect to very
accurate radial-velocity measurements, which cover at present a few years only,
our method allows us
to identify binaries with much longer periods. For stars at
$r = 100$ pc, the measuring accuracy of our method, 0.43 km/s, is still
comparable to conventional radial-velocity measurements.
Here our method has the
advantage that this accuracy is also attained for early-type stars for which
spectroscopic correlation methods have difficulties because of the small
number of spectral lines. For rather distant stars, say at $r = 1$ kpc, our
method is not very sensitive, due to a measuring error of more than 4 km/s.

For the much larger number of GC stars (Table 2),
the sensitivity of our method
is lower by a factor of about 2 with respect to the basic FK5 stars. For nearby
objects at $r = 10$ pc, however, the measuring accuracy of 77 m/s for GC stars
is still impressive. Here, our method has the advantage to reach a higher
number of objects than the published radial-velocity surveys.

Since we accept only those objects as $\Delta\mu$ binaries which have an $F$
value larger than 3.44, the velocity difference for such candidates has to be
larger than about 0.15 km/s for basic FK5 stars and 0.26 km/s for GC stars at
$r = 10$ pc, and correspondingly higher for more distant stars. For nearby
objects, say up to 25 pc, these values are still very acceptable.

\section{Interpretation of the observed velocity difference}

Our method provides primarily a qualitative indication on whether or not an
object can be classified as
$\Delta\mu$ binary or as a single-star candidate. However,
the observed velocity difference for a $\Delta\mu$ binary
obviously also contains
quantitative information on the character of this probable binary.
Binaries of medium and long periods have to be treated differently.

\subsection{Binaries with medium periods}

We call periods of the order of a few decades, say $P \sim 30$ years,
medium periods. In this case, $\Delta\mu_{FH, tot}$ corresponds to the
two-dimensional projection of the instantaneous three-dimensional velocity
$\upsilon_{ph}$ of the photo-center of the binary with respect to the
center-of-mass. For a randomly orientated velocity $\upsilon_{ph}$, we have on
average:
\begin{equation}
\upsilon_{ph} \, [{\rm km/s}] = 4.74 \,\, \Big(\frac{\pi}{4}\Big)^{-1} \,\,
\frac{\Delta\mu_{FH, tot} \, [{\rm mas/year}]}{p\,[{\rm mas}]} \,\, . 
\end{equation}
If we would prefer to use the median value of $\upsilon_{ph}$
instead of the inverse mean one,
we should replace in Eq. (40) $\pi/4 = 0.785$ by $\sqrt{3/4} = 0.866$.
In the case of the GC, we use $\Delta\mu_{0 (GC) H, tot}$ instead of
$\Delta\mu_{FH, tot}$. The velocity of the photo-center, $\upsilon_{ph}$, is
related to the instantaneous velocity of component B relative to A,
$\upsilon_{AB}$, by
\begin{equation}
\upsilon_{ph} = | B - \beta| \, \upsilon_{AB} \,\, ,
\end{equation}
where $B$
and $\beta$ are the fractions of the mass ${\cal M}$ and of the
luminosity $L$ of the secondary component B:
\begin{equation}
B
= \frac{{\cal M}_B}{{\cal M}_A + {\cal M}_B} \,\, , \,\, \beta =
\frac{L_B}{L_A + L_B} \,\,.                                             
\end{equation}
Usually, we assume that B is dark $(L_B \ll L_A, \, \beta \sim 0)$. If
desired, a finite value of $L_B$ can be taken into account later by starting an
iteration process. For an elliptic orbit of eccentricity $e$, the time average
of $\upsilon_{AB}$ is given by
\begin{equation}
\upsilon_{AB} \, [{\rm km/s}] = 4.74 \, \frac{2 \pi \, a \, [{\rm AU}]}{P \,
[{\rm years}]} \, f_{\upsilon} (e) \,\, ,                           
\end{equation}
where $a$ is the semi-major axis of the orbit of B relative to A, and
\begin{equation}
f_{\upsilon} (e) = \frac{2}{\pi} \, \vec{E}(e) \,\, .
\end{equation}
$\vec{E}(k)$ is the complete elliptic integral of the second kind.
We have $f_{\upsilon} (0) = 1$,
$f_{\upsilon} (0.5) = 0.93$, and $f_{\upsilon} (1) = 2/\pi = 0.64$. In most of
our applications we choose $e = 0.5$ as a typical value. The relation between
$a$ and the period $P$ is given by Kepler's third law:
\begin{equation}
(a \, [{\rm AU}])^3\,/\,(P \, [{\rm years}])^2 = {\cal M}_A \,
[{\cal M}_{\odot}] + {\cal M}_B \, [{\cal M}_{\odot}] \,\, .          
\end{equation}
The mass ${\cal M}_A$ can be derived from a properly chosen mass-luminosity
relation for $\beta \sim 0$. Combining Eqs. (40)\,-\,(45), we find
\begin{eqnarray}
\frac{{\cal M}_B}{({\cal M}_A + {\cal M}_B)^{2/3}} \, \sim
|B - \beta| \,
({\cal M}_A + {\cal M}_B)^{1/3} = \varphi = \nonumber\\
(2/\pi^2) \, f_{\upsilon}^{-1} (e)  \,\, (P \, [{\rm years}])^{1/3} \,
(p \, [{\rm mas}])^{-1}\nonumber\\
\, \Delta\mu_{FH, tot} \, [{\rm mas/year}] \,\, .                     
\end{eqnarray}
The quantity $\varphi$ in Eq. (46) is closely linked to the `mass function' of
spectroscopic binaries which equals $(\varphi\,{\rm sin}\,i)^3$ for $\beta =
0$, where $i$ is the inclination of the orbit. From Eq. (46) we can determine
${\cal M}_B$\,[${\cal M}_{\odot}$] (e.g. iteratively) for a given value
of $P$, say for $P = 30$ years. For $B \ll 1$,
${\cal M}_B$ scales with $P^{1/3}$. Having determined a typical value of
${\cal M}_B$ from our data, we can use again the mass-luminosity relation for
deriving $L_B$ and hence the magnitude difference $m_B - m_A$. This is also a
check on our approximation for $\beta \, (\beta \sim 0)$.

Knowing now ${\cal M}_A$ and ${\cal M}_B$  we can derive the semi-major axis
from Eq. (45). This
allows us then to predict the projected separation $\rho_{AB}$ of B and A,
and the projected distance $\rho_{ph}$ of the photo-center from the
center-of-mass:
\begin{equation}
\rho_{AB} \, [{\rm mas}] = \frac{\pi}{4} \,
f_r (e) \,\,
a \, [{\rm AU}] \,\, p \,[{\rm mas}] \,\, ,                           
\end{equation}
with
\begin{equation}
f_r (e) = 1 + \frac{1}{2} \,e^2  \,\, ,                               
\end{equation}
and
\begin{equation}
\rho_{ph} = |B - \beta| \, \rho_{AB} \,\, .              
\end{equation}
The estimated value of $\rho_{ph}$ is proportional to
the assumed period $P$,
that of $\rho_{AB}$ for $B \ll 1$ to $P^{2/3}$.
Our method allows us
therefore to predict approximate values of ${\cal M}_B, \,
\Delta m, \, \rho_{AB}$ and $\rho_{ph}$. Clearly, these estimated values have a
large statistical noise, because of the many unknowns (geometry, orbital phase,
period, eccentricity). Nevertheless, the values of $\rho_{AB}$ and $\Delta m$
are interesting for the planning of observations for direct imaging. The value
of $\rho_{ph}$ is an estimate of the (two-dimensional) cosmic error in the
HIPPARCOS position of this star. We can also calculate the expected {\it total}
variation of the radial velocity of A in time, $\Delta\upsilon_{rad}$. In a
sufficient approximation $(e = 0, \, \beta = 0)$, we find
\begin{eqnarray}
\Delta \upsilon_{rad} \, {\rm [km/s]} &  \sim   &
2.6 \,\, \Delta\upsilon_{FH, tot} \, {\rm [km/s]}    \nonumber\\
 & = & 2.6 \,\, \frac{4.74 \, \Delta\mu_{FH, tot} \, {\rm [mas/year]}}
{p \, {\rm[mas]}}
      \,\, .                                                         
\end{eqnarray}
This estimate may be helpful for
the planning of radial-velocity observations.

\subsection{Long-period binaries}

In the case of long periods,
the proper-motion differences do not allow us to
determine the orbital {\em velocity} of the photo-center. Instead we obtain the
orbital {\em acceleration}, $g_{ph}$, of the photo-center:
\begin{equation}
g_{ph, tot, 2{\rm D}} = \frac{\Delta\mu_{FH, tot}}{|T_H - T_F|} \,\, . 
\end{equation}
The components of $g_{ph}$ in $\alpha_\ast$ and $\delta$ provide immediately
the position angle of B relative to A, since the vector of $g_{ph}$ points from
A to B, if
$B - \beta > 0$. On (linear) average, we have for a
two-dimensional projection $g_{ph, tot, 2{\rm D}}$ of a three-dimensional
vector $g_{ph}$:
\begin{equation}
g_{ph, tot, 2{\rm D}} =  \frac{\pi}{4} \,\, g_{ph, tot, 3{\rm D}} \,\, .
\end{equation}                                                         
Similar to Eq. (41), we have
\begin{equation}
g_{ph, tot, 3{\rm D}} \, [{\rm mas/year}^2] =
|B - \beta| \, g_{AB} \,
[{\rm AU/year}^2] \,p \,[{\rm mas}] ,                            
\end{equation}
with
\begin{equation}
g_{AB} = a \, \left(\frac{2 \pi}{P}\right)^2 \, f_g (e)              
\end{equation}
as an average over the orbital phase, where
\begin{equation}
f_g (e) = (1 - e^2)^{-1/2} \,\, .                                    
\end{equation}
Proceeding in the same way as in Sect. 6.1, we derive
\begin{eqnarray}
\frac{{\cal M}_B}{({\cal M}_A + {\cal M}_B)^{2/3}} \sim
|B - \beta| \,
({\cal M}_A + {\cal M}_B)^{1/3} = \varphi = \nonumber\\
(1/\pi^3) \, f_g^{-1} (e) \, (P \, [{\rm
years}])^{4/3} \, (p \, [{\rm mas}])^{-1}\nonumber\\
\Delta\mu_{FH, tot} \, [{\rm mas/year}] \,
(|T_H - T_F| \, [{\rm years}])^{-1} \,\, .                          
\end{eqnarray}
We can now determine ${\cal M}_B$ from Eq. (56) for a given period
$P$, say for $P = 1000$ years.
In the case of long periods, ${\cal M}_B$ scales as
$P^{4/3}$ for
$B \ll 1$. For estimating $\rho_{AB}$ and  $\rho_{ph}$,
we use Eqs. (45), (47), and (49).
The estimated value of $\rho_{ph}$ is proportional to the square of the
the assumed period, $P^2$,
that of $\rho_{AB}$ for $B \ll 1$ to $P^{2/3}$.
The change in radial velocity
{\it during a time interval} $\Delta t$ is
expected to be (for $\beta = 0$)
\begin{eqnarray}
\Delta \upsilon_{rad} \, {\rm [km/s]} & \sim &
\Delta\upsilon_{FH, tot} \, {\rm [km/s]} \,
\frac{\Delta t \, {\rm [years]}}{|T_H - T_F| \, {\rm [years]}} \nonumber \\
& = & \frac{4.74 \, \Delta\mu_{FH, tot} \, {\rm [mas/year]}
\, \Delta t \, {\rm [years]} }
{ p \, {\rm[mas]} \,\, |T_H - T_F| \, {\rm [years]} } \, .
\end{eqnarray}
Since Eqs. (56) and (46) have a similar structure, we can easily derive the
ratio of ${\cal M}_{B, long}$ to ${\cal M}_{B, medium}$ for
$B \ll 1$ and $\beta \sim 0$:
\begin{eqnarray}
& & \left(\frac{{\cal M}_{B, long}}{{\cal M}_{B,
medium}}\right) = \nonumber\\
& & \frac{1}
{2 \pi} \, \frac{f_{\upsilon} (e)}{f_g (e)} \, \frac{(P_{long}
\,[{\rm years}])^{4/3}}{(P_{medium}\,[{\rm years}])^{1/3}\,(|T_H -
T_F|\,[{\rm years}])} \,\, .                                           
\end{eqnarray}
For $P_{long} = 1000$ years, $P_{medium} = 30$ years, $T_H - T_F = 40$ years,
and $e = 0.5$, we find a ratio of about 10.

In some cases we do not find a plausible solution for
${\cal M}_A$ and
${\cal M}_B$ from Eq.(56) or even from Eq.(46): For $\beta=0$, we may obtain
that the mass ${\cal M}_B$ of the secondary is
larger than or comparable to the mass
${\cal M}_A$ of the primary, which is not very probable. If we allow for
$\beta\not=0$ and assume that both components are main-sequence stars, we find
a solution for
${\cal M}_A$ and ${\cal M}_B$ only if
$(\varphi / {\cal M}_{A,{\beta=0}}^{1/3})
\,\, ^<_{\sim}\,\,  0.3$\,.
Of course, we may then change the period $P$ from
its assumed standard value to such a lower value that the estimates
for ${\cal M}_A$
and ${\cal M}_B$ become now reasonable.

In Sects. 6.1 and 6.2, we have neglected the difference between the expectation
value $\langle q \rangle$ of a quantity $q$
and $1/\langle 1/q \rangle$,
and between $\langle q \rangle ^n$ and $\langle q^n \rangle $.
For most quantities, this difference is smaller than the inherent
`noise' in the expectation value caused by the unknown orbital phase of the
star and by the unknown spatial orientation of its orbit, which are treated
both in a statistical way only.

\section{A few examples}

In order to illustrate our method we provide in
Tables 3 and 4 a few examples.
Individual comments on these stars are given in the following subsections. A
discussion of other FK and GC objects will follow in subsequent papers.

\begin{table*}[th]
\caption{
Some examples of $\Delta\mu$ binaries
($\gamma$ UMa, $\varepsilon$ Eri, $\iota$ Vir)
and of single-star candidates (47 UMa, $\delta$ Pav)
}
\begin{center}
\begin{tabular}{llrrrrrrrrrr}
\hline\\[-1.5ex]
&  HIP-No.: & 58001 & & 16537 & & 69701 & & 53721 & & 99240 &\\
&  FK-No.:  &   447 & &   127 & &   525 & & 1282 & &   754 &\\
&  Name:    & $\gamma$ UMa & & $\varepsilon$ Eri & & $\iota$ Vir & & 47 UMa & &
$\delta$ Pav &\\
Quantity & Unit &  & & & & & & & & &\\[0.7ex]\hline\\[-2ex]
$p_H$ & [mas]  & 38.99 & $\pm$ 0.68 & 310.75 & $\pm$ 0.85 & 46.74 & $\pm$ 0.87
& 71.04 & $\pm$ 0.66 & 163.73 & $\pm$ 0.65\\
$r$   & [pc] & 25.65 & $\pm$ 0.45 & 3.22 & $\pm$ 0.01 & 21.39 & $\pm$ 0.40
& 14.08 & $\pm$ 0.13 & 6.11 & $\pm$ 0.02\\[1.5ex]
$\varepsilon_{\mu, H, \alpha\ast}$ & [mas/yr] & & $\pm$ 0.48 & & $\pm$ 0.98 & &
$\pm$ 0.91 & & $\pm$ 0.58 & & $\pm$ 0.55\\
$\varepsilon_{\mu, H, \delta}$ & [mas/yr] & & $\pm$ 0.50 & & $\pm$ 0.91 & &
$\pm$ 0.68 & & $\pm$ 0.54 & & $\pm$ 0.47\\
$\varepsilon_{\mu, F, \alpha\ast}$ & [mas/yr] & & $\pm$ 0.47 & & $\pm$ 0.45 & &
$\pm$ 0.53 & & $\pm$ 0.62 & & $\pm$ 0.98\\
$\varepsilon_{\mu, F, \delta}$ & [mas/yr] & & $\pm$ 0.44 & & $\pm$ 0.52 & &
$\pm$ 0.59 & & $\pm$ 0.71 & & $\pm$ 1.06\\
$\varepsilon_{\mu, 0, \alpha\ast}$ & [mas/yr] & & $\pm$ 0.35 & & $\pm$ 0.30 & &
$\pm$ 0.33 & & $\pm$ 0.45 & & $\pm$ 0.80\\
$\varepsilon_{\mu, 0, \delta}$ & [mas/yr] & & $\pm$ 0.29 & & $\pm$ 0.29 & &
$\pm$ 0.32 & & $\pm$ 0.48 & & $\pm$ 0.78\\[1.5ex]
$\varepsilon_{\mu, 0\,(GC), \alpha\ast}$ & [mas/yr] & & $\pm$ 0.49 & & $\pm$
0.33 & & $\pm$ 0.37 & & $\pm$ 0.56 & & $\pm$ 0.82\\
$\varepsilon_{\mu, 0\,(GC), \delta}$ & [mas/yr] & & $\pm$ 0.40 & & $\pm$ 0.32 &
& $\pm$ 0.36 & & $\pm$ 0.53 & & $\pm$ 0.84\\[1.5ex]
$\Delta\mu_{FH, \alpha\ast}$ & [mas/yr] & -- 12.94 & $\pm$ 0.67 & + 1.70 &
$\pm$ 1.08 & + 21.05 & $\pm$ 1.05 & -- 1.73 & $\pm$ 0.85 & + 0.71 & $\pm$
1.12\\
$\Delta\mu_{FH, \delta}$ & [mas/yr] & --\hphantom{1} 1.36 & $\pm$ 0.67 & + 3.31
& $\pm$ 1.05 & -- 12.21 & $\pm$ 0.90 & + 0.85 & $\pm$ 0.89 & -- 0.81 & $\pm$
1.16\\
$\Delta\mu_{FH, tot}$ & [mas/yr] & 13.01 & $\pm$ 0.69 & 3.72 & $\pm$ 1.16 &
24.33 & $\pm$ 1.10 & 1.93 & $\pm$ 0.89 & 1.08 & $\pm$ 1.17\\[1.5ex]
$\Delta\mu_{0H, \alpha\ast}$ & [mas/yr] & -- 12.63 & $\pm$ 0.59 & + 1.76 &
$\pm$ 1.02 & + 10.52 & $\pm$ 0.97 & -- 1.10 & $\pm$ 0.73 & + 0.60 & $\pm$
0.97\\
$\Delta\mu_{0H, \delta\ast}$ & [mas/yr] & --\hphantom{1} 0.81 & $\pm$ 0.58 & +
2.49 & $\pm$ 0.96 & -- 14.29 & $\pm$ 0.75 & -- 0.31 & $\pm$ 0.72 & -- 1.16 &
$\pm$ 0.91\\
$\Delta\mu_{0H, tot}$ & [mas/yr] & 12.66 & $\pm$ 0.60 & 3.05 & $\pm$ 1.11 &
17.74 & $\pm$ 0.95 & 1.14 & $\pm$ 0.71 & 1.31 & $\pm$ 0.95\\[1.5ex]
$\Delta\mu_{0F, \alpha\ast}$ & [mas/yr] & + 0.31 & $\pm$ 0.59 & + 0.06 &
$\pm$ 0.54 & -- 10.53 & $\pm$ 0.62 & + 0.63 & $\pm$ 0.77 & -- 0.11 & $\pm$
1.27\\
$\Delta\mu_{0F, \delta}$ & [mas/yr] & + 0.55 & $\pm$ 0.53 & -- 0.82 & $\pm$
0.60 & --\hphantom{1} 2.08 & $\pm$ 0.67 & -- 1.16 & $\pm$ 0.86 & -- 0.35 &
$\pm$ 1.32\\
$\Delta\mu_{0F, tot}$ & [mas/yr] & 0.63 & $\pm$ 0.55 & 0.82 & $\pm$ 0.60 &
10.73 & $\pm$ 0.62 & 1.32 & $\pm$ 0.84 & 0.37 & $\pm$ 1.32\\[1.5ex]
$\Delta\mu_{0\,(GC)\,H, \alpha\ast}$ & [mas/yr] & -- 13.42 & $\pm$ 0.69 & +
1.40 & $\pm$ 1.03 & + 16.24 & $\pm$ 0.98 & -- 0.58 & $\pm$ 0.81 & + 0.31 &
$\pm$ 0.99\\
$\Delta\mu_{0\,(GC)\,H, \delta}$ & [mas/yr] & -- \hphantom{1}1.92 & $\pm$ 0.64
& + 3.82 & $\pm$ 0.96 & -- 12.62 & $\pm$ 0.77 & + 0.44 & $\pm$ 0.76 & + 0.31 &
$\pm$ 0.96\\
$\Delta\mu_{0\,(GC)\,H, tot}$ & [mas/yr] & 13.56 & $\pm$ 0.72 & 4.07 & $\pm$
1.06 & 20.57 & $\pm$ 1.01 & 0.73 & $\pm$ 0.83 & 0.44 & $\pm$ 0.95\\[1.5ex]
$F_{FH}$ & [number] & 19.78 & & 3.26 & & 22.09 & & 2.18 & & 0.92 &\\
$F_{0H}$ & [number] & 22.68 & & 2.78 & & 19.82 & & 1.62 & & 1.38 &\\
$F_{0F}$ & [number] & 1.18 & & 1.39 & & 17.19 & & 1.59 & & 0.28 &\\[1.5ex]
$F_{0\,(GC)\,H}$ & [number] & 19.85 & & 3.96 & & 20.60 & & 0.88 & & 0.46
&\\[1.5ex]
$\Delta\upsilon_{FH, tot}$ & [km/s] & 1.58 & $\pm$ 0.08 & 0.057 & $\pm$ 0.018 &
2.47 & $\pm$ 0.11 & 0.129 & $\pm$ 0.059 & 0.031 & $\pm$ 0.034\\
$\Delta\upsilon_{0H, tot}$ & [km/s] & 1.54 & $\pm$ 0.07 & 0.047 & $\pm$ 0.017 &
1.80 & $\pm$ 0.10 & 0.076 & $\pm$ 0.047 & 0.038 & $\pm$ 0.028\\
$\Delta\upsilon_{0F, tot}$ & [km/s] & 0.08 & $\pm$ 0.07 & 0.013 & $\pm$ 0.009 &
1.09 & $\pm$ 0.06 & 0.088 & $\pm$ 0.056 & 0.011 & $\pm$ 0.038\\[1.5ex]
$\Delta\upsilon_{0\,(GC)\,H, tot}$ & [km/s] & 1.65 & $\pm$ 0.09 & 0.062 &
$\pm$ 0.016 & 2.09 & $\pm$ 0.10 & 0.049 & $\pm$ 0.055 & 0.013 & $\pm$
0.028\\[0.5ex]\hline
\end{tabular}
\end{center}
\end{table*}

\subsection{\object{HIP 58001} = $\gamma$ UMa}

This is a very good example for a new binary detected by our $\Delta\mu$
method. The high values of $F_{FH}, \, F_{0H}$, and $F_{0 (GC) H}$ (of the
order of 20) leave no doubt on the binary nature of $\gamma$ UMa. The FK5 and
GC values are in perfect agreement. Since $F_{0F}$ is rather small (indicating
a good agreement between the `mean' proper motions $\mu_F$ and $\mu_0$),
$\gamma$ UMa is most probably a binary with a medium period $P$ of a few
decades. Earlier, not very detailed reports on a variability of the radial
velocity of $\gamma$ UMa have not been confirmed by more recent observations,
according to Hubrig \& Mathys (1994). Since $\gamma$ UMa A has a large
rotational velocity ($\upsilon_{rot}\, {\rm sin}\,i = 165$ km/s, Abt \&
Morrell 1995), accurate radial-velocity measurements are difficult.
The expected total variation of the radial velocity, about 4 km/s, is so small
that it is not astonishing that $\gamma$ UMa has not been detected as a
spectroscopic binary up to now.

For an assumed medium period of $P =$  30 years,
we predict in Table 4 a separation
between the two components of $\gamma$ UMa of about
$0\farcs5$ and a magnitude
difference in V of nearly $10^m$. It will certainly be very difficult to
observe the secondary component, probably a late dwarf or a white dwarf, by
direct methods. A very long period is improbable for $\gamma$ UMa, since this
would lead to ${\cal M}_B > {\cal M}_A$
in our statistical estimate of ${\cal M}_B$ for $\beta = 0$. This
confirms our earlier conclusion on
$P$ based on the small value of $F_{OH}$.

$\gamma$ UMa is one of the members of the Ursa Major Star Cluster. By using the
quality of the convergence of the FK4 proper motions of 6 members of the UMa
cluster, Wielen (1978a, b) has shown that the internal velocity dispersion of
this cluster is as small as 0.1 km/s, corresponding to 1 mas/year at the
distance of the cluster ($r \sim 25$ pc). The FK4 proper motions are obviously
not affected by cosmic errors and represent already very closely `mean'
proper motions. The fact that the FK4 proper motion gave a cluster parallax for
$\gamma$ UMa (39.3\,$\pm$\,0.3 mas) which is in perfect agreement with the
HIPPARCOS trigonometric parallax (39.0\,$\pm$\,0.7 mas) is a very strong
indication that the cosmic error in the HIPPARCOS proper motion of $\gamma$ UMa
is due to a binary motion
and is not artificially caused by a wrong estimate of
the measuring errors in the ground-based proper
motions. In contrast, a
cluster parallax of $\gamma$ UMa derived by using the HIPPARCOS
proper motion would be larger than the HIPPARCOS trigonometric parallax $p_H$
by about +\,5 mas or 8 mean errors of $p_H$. By a mere accident the cosmic
error (about 13 mas/year) in the HIPPARCOS proper motion of $\gamma$
UMa points nearly exactly towards the convergent point of the UMa cluster.
There is certainly no physical reason behind this coincidence. While $\zeta$
UMa is a well-known multiple system, none of the remaining 4 FK members of the
UMa cluster are $\Delta\mu$ binaries. However, it is also true that none of
these 4 FK stars is qualified as a single-star candidate.

\subsection{\object{HIP 16537} = $\varepsilon$ Eri}

This star is an example for a $\Delta\mu$ binary at the verge of
detectability and significance.
The value
$F_{0 (GC) H}=3.96$ is larger than $F_{lim,b}=3.44$, while
$F_{FH}=3.26$ is slightly smaller than our adopted limit.
We measure a velocity discrepancy
$\Delta\upsilon_{FH, tot}$ of 57 m/s. The fact that $\Delta\mu_{FH} \sim
\Delta\mu_{0H}$ while $\Delta\mu_{0F} \sim 0$, favours an orbital period of
medium length.

Accurate radial-velocity measurements by Walker et al. (1995) suggest a period
of $P = 9.88$ years and a radial-velocity amplitude of about 15 m/s. The
significance of this result is disputed by the authors themselves (Walker et
al. 1995) and by others (see Greaves et al. 1998). If the radial-velocity
amplitude of 15 m/s and our value of $\Delta\upsilon_{FH, tot}$ (57 m/s) are
correct, then the orbital inclination $i$ must be quite small, of the order of
10\,-\,20$^\circ$. This would be in agreement with the orientation of the
stellar pole of $\varepsilon$ Eri which corresponds to $i \sim 30 \pm
15^\circ$, as deduced by Saar \& Osten (1997). Using $i = 15^\circ$ and the
data given in Table 4, the mass of the secondary component of $\varepsilon$
Eri would be about 4 Jupiter masses, corresponding to a massive planet or a
low-mass brown dwarf. The semi-major axis of the photo-center of $\varepsilon$
Eri would be 6 mas, and the semi-major axis of the orbit
of the planet would be about 4 AU or 1{\farcs}4\,.

Blazit et al. (1977) published a speckle measurement of $\varepsilon$ Eri,
giving a binary separation of $48 \pm 5$ mas. Later speckle observations did
not resolve the star (with upper limits of about 30\,-\,35 mas). In any case,
such a close component would not be detected by our $\Delta\mu$ method, since
its orbital period (about 20 days) would be much smaller than the 3 years over
which the HIPPARCOS proper motion of $\varepsilon$ Eri is averaged.

Gatewood (private communication in 1998) found
by using the Multichannel Astrometric Photometer (MAP)
that any deviation of
$\varepsilon$ Eri
from a linear motion is smaller than about 1.4 mas over a period
of time of about eight years.
This is in contradiction with our estimate of $a_{ph}\sim 6$ mas, based on
$\Delta\mu \sim$ 4 mas/year.
Even for a much longer period $P$,
our value of $\Delta\mu$ would imply larger
deviations than the upper limit claimed by Gatewood. The reason for this
discrepancy is unclear.

Greaves et al. (1998) detected a dust ring around $\varepsilon$ Eri by
measuring the dust emission at $\lambda = 0.85$ mm. The asymmetries and
substructures within this ring can hardly be due the potential planet discussed
above, since the ring has a radius of about 30 AU, much larger than the
semi-major axis of the planet discussed here ($a \sim 4$ AU). As noted by
Greaves et al. (1998) the $\varepsilon$ Eri ring system is roughly circular and
hence appears close to face-on, in agreement with the small inclination derived
from our comparison of $\Delta\upsilon_{FH, tot}$ with the radial-velocity
amplitude, if we assume co-planarity of the planetary orbit and of the ring.

\subsection{\object{HIP 69701} = $\iota$ Vir}

This star is an example for a probable long-period binary. At least the
proper-motion differences in $\alpha_\ast$ are consistent with
our criterion for a long period, given
by Eq. (36). The proper-motion differences in $\delta$,
however, do not obey Eq. (36). This may indicate that the orbital period of
$\iota$ Vir is not very long, perhaps of the order of 200 years.

The amplitude of the orbital motion of the photo-center of $\iota$ Vir amounts
to a few hundred mas at least, based on our estimate of
$\rho_{ph} \sim 0{\farcs}5$ for $P = 200$ years.
This amplitude is confirmed by the results of modern
meridian-circle observations used in the construction of the FK5
(Schwan, private communication in 1990).
The unusual behaviour of $\iota$ Vir has also been noted by
Morrison et al. (1990) in a comparison of recent meridian-circle positions with
the FK5 prediction. Our $\Delta\mu$ method fully confirms the earlier
suggestion that $\iota$ Vir is a double star.

\subsection{\object{HIP 53721} = 47 UMa}

From radial-velocity measurements, Butler \& Marcy (1996) have detected a
planet orbiting 47 UMa. The orbital period is 3.0 years and the radial-velocity
amplitude is 48 m/s. The direct positional measurements by HIPPARCOS are not
accurate enough to confirm the orbital motion of 47 UMa due to this planet
(Perryman et al. 1996). Our $\Delta\mu$ method is also not able to detect such
a planet since its orbital
period is too short (the HIPPARCOS proper motion is averaged
over 3 years) and the velocity amplitude is rather small (our measuring error
of $\Delta\upsilon$ is of the order of 60 m/s for 47 UMa). Our $\Delta\mu$
method indicates strongly, however, that 47 UMa is otherwise a single-star
candidate, i.e. that no massive companion of 47 UMa exists, since all the $F$
values of 47 UMa are below our limit of 2.49. The largest velocity discrepancy
is $\Delta\upsilon_{FH, tot}$ = 129 $\pm$ 62 m/s, i.e. not significantly
different from zero. This is confirmed by the radial-velocity observations
which do not show any systematic trend in time (beside the 3-years period). In
summary, 47 UMa seems to be a {\it single star}, but with at least one planet.

\subsection{\object{HIP 99240} = $\delta$ Pav}

This is an example for a good single-star candidate. All its $F$ values are
rather small, below 1.4 in each case.
The measuring error of the total velocity
discrepancy $\Delta\upsilon$ is about 30 m/s, and hence comparable to that of
modern radial-velocity measurements. The published radial velocities do not
show any significant variations in time. This excludes massive secondaries with
periods below the period limit of our $\Delta\mu$ method (i.e. a few years),
if the orbit is not nearly face-on. A low-mass secondary with a short period
cannot be ruled out. A very long-period companion can be probably
excluded, because its separation of many seconds of arc would have allowed its
visual detection, if the magnitude difference is not extremely large.

\begin{table}[th]
\caption{
Estimated values of the masses, magnitude differences,
and separations for three examples of $\Delta$$\mu$ binaries
}
\begin{center}
\begin{tabular}{llrrr}
\hline\\[-1.5ex]
&  HIP-No.: & 58001 & 16537 & 69701\\
&  FK-No.:  &   447 &   127 &   525\\
&  Name:    & $\gamma$ UMa & $\varepsilon$ Eri & $\iota$ Vir\\
Quantity & Unit &  &  &\\[0.7ex]\hline\\[-2ex]
$m_{V, tot}$ & [mag] & 2.41 & 3.72 & 4.07\\
$M_{V, tot}$ & [mag] & + 0.36 & + 6.18 & + 2.42\\
${\cal M}_A \, (\beta$=0)  & [${\cal M}_{\odot}$] & 3.07 & 0.85 & 1.64\\[2ex]
\multicolumn{5}{c}{Standard medium period $P = 30$ years and $e = 0.5$
assumed:}\\[1.5ex]
$\varphi$ & [${\cal M}_{\odot}^{1/3}$] & 0.23 & 0.0081 & 0.35\\
${\cal M}_A$ & [${\cal M}_{\odot}$] & 3.07 & 0.85 & 1.64\\
${\cal M}_B$ & [${\cal M}_{\odot}$] & 0.53 & 0.0073 & 0.61\\
$\Delta m _{AB}$ & [mag] & 9.4 & large & 6.2\\
$\rho_{AB}$ & [mas] & 518 & 2560 & 531\\
$\rho_{ph}$ & [mas] & 76 & 22 & 141\\
$\Delta\upsilon_{rad}$ & [km/s] & 4.1 & 0.15 & 6.4\\[2ex]
\multicolumn{5}{c}{Standard long period $P = 1000$ years and $e = 0.5$
assumed:}\\[1.5ex]
$\varphi$ & [${\cal M}_{\odot}^{1/3}$] & 1.78 & 0.062 & 2.80\\
${\cal M}_A\, (\beta$=0) & [${\cal M}_{\odot}$] & (3.07) & 0.85 &  (1.64)\\
${\cal M}_B\, (\beta$=0) & [${\cal M}_{\odot}$] & (9.70) & 0.06 & (24.91)\\
$\Delta m_{AB}$ & [mag] & & large &\\
$\rho_{AB}$ & [mas] & & 27033 &\\
$\rho_{ph}$ & [mas] & & 1754 &\\
$\Delta\upsilon_{rad}$ & [km/s] & ${^{(**)}}3.0$  &
                     ${^{(*)}}0.009$ & ${^{(**)}}4.8$ \\[2ex]
\multicolumn{5}{c}{A plausible individual solution:}\\[1.5ex]
$P$ & [years] & 30.0 & obs:\,9.88 & 200\\
$e$ & [number] &  0.5 & obs:\,0.0 & 0.5\\
$i$ & [$^{\circ}$] & stat. & 15 & stat.\\
$\varphi$ & [${\cal M}_{\odot}^{1/3}$] & (see & & 0.32\\
${\cal M}_A$ & [${\cal M}_{\odot}$] & above) & 0.85 & 1.64\\
${\cal M}_B$ & [${\cal M}_{\odot}$] & & 0.0038 & 0.55\\
$\Delta m_{AB}$ & [mag] & & large & 6.9\\
$\rho_{AB}$ & [mas] & & $a_{AB}$:\,1355 & 1865\\
$\rho_{ph}$ & [mas] & & $a_{ph}$:\,6 & 466\\
$\Delta\upsilon_{rad}$ & [km/s] & & obs:\,0.030 & ${^{(**)}}4.8$
\\[0.5ex]\hline
\end{tabular}\\
\end{center}
\hspace*{0.4cm}obs: \,  Value taken from other observations.\\
\hspace*{0.4cm}stat: \, Statistical estimate.\\
\hspace*{0.4cm}${^{(*)}}$: for $\Delta t = 10$ years; \,\,\,
                        ${^{(**)}}$: for $\Delta t = 100$ years.
\end{table}

\section{Conclusions and outlook}

We have presented a new method to detect double stars:
Our $\Delta\mu$ method is
based on a comparison of the HIPPARCOS proper motions with ground-based data
provided e.g. by the FK5 or the GC. We therefore call these newly detected
double stars '$\Delta\mu$ binaries'.

If the HIPPARCOS proper motion of a star is in good agreement with the
ground-based data and if no other indications for a binary nature of the object
exist, then we classify the star as a 'single-star candidate'.

For nearby stars, our $\Delta\mu$ method is very sensitive: At a distance of
e.g. 10 pc, our measuring accuracy of orbital velocities is of the order of 50
m/s for basic FK5 stars, and of 80 m/s for many GC stars.

For the detected $\Delta\mu$ binaries we obtain statistical estimates for the
separation $\rho$
and for the magnitude difference $\Delta m$
between the components, based on
adopted orbital periods $P$.

Our statistical estimates on the parameters of a $\Delta\mu$ binary would be
significantly improved if accurate radial-velocity
measurements would at least provide
the acceleration component $g_{rad, A}$ from the linear change in time of the
radial velocity $v_{rad}$ of the double-star component A. Of course, such a
change would also be a desirable confirmation of the double-star nature of the
object.

The sensitivity of and the information provided by the $\Delta\mu$ method would
be very much increased if a future astrometric satellite (like the proposed
projects GAIA, DIVA, SIM) would reobserve the HIPPARCOS stars. In such a
case, a second `instantaneous' proper motion at the new epoch and an additional
`intermediate' proper motion (based on the two instantaneous positions) would
be available. However, even in this case, the long-term proper motions derived
by using ground-based data would remain useful for our $\Delta\mu$ method,
especially for binaries among the bright stars with longer orbital periods.

In Tables 3 and 4 we give a few examples of $\Delta\mu$ binaries and of
single-star candidates, in order to illustrate our method.

In subsequent papers we shall present the individual results of our $\Delta\mu$
method for all the appropriate FK5 and GC stars. The number of newly detected
$\Delta\mu$ binaries among these stars is more than one thousand.
The fraction is highest among the 1535 basic FK5 stars:
about 10 percent of them are
newly discovered $\Delta \mu$ binaries.

\begin{acknowledgements}
We thank J. Schubart for supplying the formulae for $f_r(e),
f_\upsilon(e)$, and $f_g(e)$,
and B. Fuchs for checking our adopted projection factors.
\end{acknowledgements}


\end{document}